\documentclass[pra,twocolumn,showpacs,preprintnumbers,amsmath,amssymb,superscriptaddress]{revtex4}

\usepackage{graphicx}
\usepackage{dcolumn}
\usepackage{bm}
\usepackage{amsmath}

\def \ed {\end{document}}
\def\Fbox#1{\vskip1ex\hbox to 8.5cm{\hfil\fboxsep0.3cm\fbox{%
  \parbox{8.0cm}{#1}}\hfil}\vskip1ex\noindent}  

\def\be{\begin{equation}}\def\ee{\end{equation}}
\def\bea{\begin{eqnarray}}\def\eea{\end{eqnarray}}
\def\bse{\begin{subequations}}\def\ese{\end{subequations}}
\newcommand{\BE}[1]{\begin{equation}\label{#1}}
\newcommand{\BEA}[1]{\begin{eqnarray}\label{#1}}
\newcommand{\BSE}[1]{\begin{subequations}\label{#1}}

\let \= \equiv \let\*\cdot \let\~\widetilde \let\^\widehat \let\-\overline

\def\<{\left\langle}    \def\>{\right\rangle}
\def\({\left(}          \def\){\right)}
 \def \[ {\left [} \def \] {\right ]}

\makeatletter
\AtBeginDocument{\@ifpackageloaded{natbib}{\ifNAT@numbers\if@filesw\immediate\write\@auxout{\string\global\string\NAT@numberstrue}\fi\fi}{}}
\makeatother

\begin{document}

\preprint{APS/123-QED}

\title{Direct and inverse cascades of spin-wave turbulence in spin-1 ferromagnetic spinor Bose-Einstein condensates} 

\author{Kazuya Fujimoto}
\affiliation{Department of Physics, Osaka City University, Sumiyoshi-ku, Osaka 558-8585, Japan}

\author{Makoto Tsubota}
\affiliation{Department of Physics, Osaka City University, Sumiyoshi-ku, Osaka 558-8585, Japan}
\affiliation{The OCU Advanced Research Institute for Natural Science and Technology (OCARINA), Osaka City University, Sumiyoshi-ku, Osaka 558-8585, Japan}

\date{\today}

\begin{abstract}
We theoretically and numerically study spin-wave turbulence in spin-1 ferromagnetic spinor Bose-Einstein condensates, finding direct and inverse cascades with power-law behavior. To derive these power exponents analytically, the conventional weak wave turbulence theory is applied to the spin-1 spinor Gross-Pitaevskii equation. Thus, we obtain the $-7/3$ and $-5/3$ power laws in the transverse spin correlation function for the direct and inverse cascades, respectively. To confirm these power laws, numerical calculations are performed that obtain results consistent with these power laws. 
\end{abstract}

\pacs{03.75.Kk,05.45.-a}

\maketitle

\section{Introduction}
Ultracold atomic gases give us an excellent stage to study various nonequilibrium phenomena such as dynamical phase transitions, thermalization in  isolated quantum systems \cite{thermal0}, and quantum hydrodynamics \cite{qhydor0}. Actually, experimental studies of the Kibble-Zurek mechanism \cite{KZ1,KZ2,KZ3,KZ4}, thermalization in isolated one-dimensional systems \cite{thermal1,thermal2}, and quantum turbulence (QT) \cite{Henn09, Reeves12,Neely13,Kwon14} are widely performed, revealing novel nonequilibrium phenomena. 
 
Quantum turbulence has recently become an active area in the field of atomic Bose-Einstein condensates (BECs). 
Such systems can provide access to topics not  addressable in superfluid helium systems \cite{Hal,Vinen,Skrbek12}, e.g., QT in  two-dimensional systems and  in multicomponent BECs, thereby opening new directions in QT research. At present, in experimental studies of one-component atomic BECs, both three- and two-dimensional QT with many quantized vortices can be generated, and have been investigated in terms of anomalous expansion \cite{Henn09}, vortex distribution \cite{Reeves12,Neely13}, and vortex decay  \cite{Kwon14}. Similarly, theoretical and numerical studies also address such  QT, providing discussion of features characteristic of quantized vortices such as the Kolmogorov $-5/3$ power law \cite{Parker,KT07}, direct and inverse cascades \cite{Gou09,Num10,Reeves13,Billam14}, nonthermal fixed points \cite{Bor11,Bor12,Sch12}, the probability density function for superfuild velocity \cite{White10}, vortex distributions \cite{White12,Bradley12,Simula14}, and vortex decay  \cite{Stagg15}. Furthermore, QT in multicomponent BECs has also been theoretically and numerically studied, giving nontrivial results for correlation functions of the wave function, the spin-density vector, and the velocity field \cite{Berl06,Take10,Karl13,Koba14,Vill14,FT14}. 

Apart from such hydrodynamic turbulence with many quantized vortices, there is another kind of turbulence: weak wave turbulence (WWT) dominated by weakly interacting waves \cite{wt1,wt2}. Ultracold gas is a superclean system with many kinds of quantum fluids such as binary BECs \cite{KTU}, spinor BECs \cite{KU,Sta}, and dipolar BECs \cite{dipole1,dipole2}, and it is possible to observe  local quantities, e.g., the density profile and the spin-density vector, so ultracold gases is one of the most suitable systems for investigating nonlinear and nonequilibrium wave dynamics such as WWT. 

However, there have been only a few  WWT studies in atomic BECs \cite{Lvov2003,Naz1,Naz2,Zakharov05,Proment09,FT15} where the Kolmogorov-Zakharov (KZ) spectrum \cite{wt1,wt2} has been discussed analytically and numerically in one-component atomic BECs. 
For instance, Proment $et$ $al.$ \cite{Proment09} investigated the connection between  three- and four-wave turbulence from the viewpoint of the growth of the condensate, studying the change of power exponents in a correlation function for a macroscopic wave function. Our recent work \cite{FT15} focused on an observable density profile, and we discussed the experimental observation of the direct cascade related to the KZ spectrum. 

In this paper, we focus on WWT in a spin-1 ferromagnetic spinor BEC dominated by spin waves, which we call spin-wave turbulence (SWT). Applying  WWT theory \cite{wt1,wt2} to a spin-1 spinor Gross-Pitaevskii (GP) equation, we analytically find the $-7/3$ and $-5/3$ power laws in the transverse spin correlation function for direct and inverse cascades, respectively. To confirm these laws, we perform numerical calculations, obtaining  results consistent with these power exponents. In this paper, we report  details of these analytical and numerical results.

Before proceeding to the next section, we note that such a nonequilibrium phenomenon for the spin waves has already been studied in solid state physics \cite{Zak74,Bry88,Aze91}. Previous experiments have focused on spin waves in yttrium iron garnet (YIG), where parametric instability, chaos, SWT, etc. were investigated. The distribution of  spin waves in SWT, as far as we know, has been addressed only in a single theoretical study \cite{Lut}. 
In magnetic materials such as YIG, the equation of motion for the spin wave is very complicated because many effects such as anisotropic fields, dipolar interactions, and thermal baths exist \cite{Zak74,Bry88}. Thus, the dynamics of spin waves observed in  previous studies is difficult to understand.  
Also, the experimental methods are based on the absorption of the pump energy and thus  do not access the spatial distribution of the spin. 

In distinction from solid state physics, the atomic BEC is a superclean system without any impurities and is highly controllable, so that it is a suitable system for studying  spin-wave dynamics. Actually, the dispersion relation and dynamics of spin waves in the spinor BEC have been experimentally and theoretically studied \cite{Mar14,Sai15,Tha14,kun15,Kaw07,Eto14}. Furthermore, because the spatial profile of the spin-density vector is observable \cite{KZ1,Sta,Ven},  it is possible to investigate the distribution of spin waves and the spin correlation function, which is advantageous in research on spin-wave dynamics. Therefore,  atomic BECs provide an excellent stage for studying spin waves. 

Finally, let us comment on the difference between this work and our previous one \cite{FT14}. The previous work focused on spin turbulence that is strongly excited by the counterflow instability, where the spin-density vector points in various directions and domain walls and vortices are nucleated. In contrast to this spin turbulence, in this paper we address weakly excited turbulence with a small-amplitude spin wave. Thus  SWT is another type of turbulence that is distinct from the spin turbulence addressed in our previous paper. 

This article is organized as follows. Section II describes the formulation, where we introduce the spin-1 spinor GP equation and spin hydrodynamic equation. In Sec. III, we apply WWT theory to the GP equation, deriving the power laws for the transverse spin correlation function. In Sec. IV, we show our numerical results for SWT. The experimental observation of these power laws is commented upon in Sec. V. Finally, we summarize this work in Sec. VI. 

\section{Formulation}
Our model for studying SWT in spinor BECs is introduced. 
First, we explain the spin-1 spinor GP equation describing the dynamics of the spin-1 spinor BECs at zero temperature. 
Second, we give the spin hydrodynamic equation equivalent to the GP equation, which is useful for the application of WWT theory. 

\subsection{Spin-1 spinor GP equation}
We consider a uniform system without a magnetic field  comprised of $N$-particle spin-1 bosons at zero temperature, which is well described by the macroscopic wave functions $\psi _m$ $(m=1,0,-1)$ obeying the spin-1 spinor GP equations \cite{Ohmi98,Ho98}: 
\begin{eqnarray}
i\hbar \frac{\partial}{\partial t} \psi _m  =  -\frac{\hbar^2}{2M} \nabla^2 \psi _m + c_0 \rho \psi _m + c_1 \bm{F}\cdot(\hat{\bm{F}})_{mn} \psi _n.  \label{GP}
\end{eqnarray}
In this paper, Roman indices appear twice,  being summed over $-1$, 0, and 1, and,  in the same way, Greek indices are summed over $x$, $y$, and $z$. The total density and the spin-density vector are given by $\rho =  \psi _m ^{*} \psi _m $ and  $F_{\mu} =  \psi _{m}^{*} ({\hat F}_{\mu})_{mn} \psi _{n}$ ($\mu = x, y, z$), respectively, where $({\hat F}_{\mu})_{mn}$ are the spin-1 matrices. The parameters $M$, $c_{0}$, and $c_{1}$ refer to the particle mass and the spin-independent and spin-dependent interactions, respectively. The sign of the coefficients $c_{1}$ plays an important role in the spin dynamics. In this paper, we consider only the ferromagnetic interaction $c_{1} < 0$.

\subsection{Spin hydrodynamic equation}
To prepare for the following sections, we introduce the spin hydrodynamic equation derived from Eq. (\ref{GP}). This kind of equation has been discussed in some papers \cite{Lam,Bar,KK,YU12}. Here, we use the hydrodynamic equation derived by Yukawa and Ueda \cite{YU12}. This equation is composed of the equations of the total density $\rho$, the superfluid velocity $\bm{v}$, the spin vector $f_{\mu} = F_{\mu}/\rho$, and the nematic tensor $n_{\mu \nu} = \psi _{m}^{*} (\hat{N}_{\mu \nu})_{mn} \psi _{n} /\rho,$ with $(\hat{N}_{\mu \nu})_{mn}  =  [ (\hat{F}_{\mu})_{ml} (\hat{F}_{\nu})_{ln} + (\hat{F}_{\nu})_{ml} (\hat{F}_{\mu})_{ln} ]/2$ and $\bm{v} = \hbar(\psi_m^*\bm{\nabla} \psi_m - \psi_m\bm{\nabla} \psi_m^*)/2Mi$.

In this paper, we consider the ferromagnetic interaction, so that the macroscopic wave functions can be assumed to be expressed by the fully spin-polarized state. Hence, the spin vector and the nematic tensor satisfy the relation
\begin{eqnarray}
n_{\mu\nu} = \frac{\delta _{\mu \nu} + f_{\mu} f_{\nu}}{2}.
\end{eqnarray}
This assumption is valid if the interaction $c_{1}$ is negative and the excitation to the system is not very strong. 
Then, by eliminating the nematic tensor in the hydrodynamic equations of \cite{YU12}, we obtain
\begin{eqnarray}
\frac{\partial}{\partial t} \rho +  \bm{\nabla} \cdot \rho \bm{v} = 0, \label{sh0}
\end{eqnarray}
\begin{eqnarray}
\frac{\partial}{\partial t} \rho f_{\mu} +  \bm{\nabla} \cdot \rho \bm{v}_{\mu} = 0, \label{sh1}
\end{eqnarray}
\begin{eqnarray}
\bm{v}_{\mu} = f_{\mu} \bm{v} - \frac{\hbar}{2M} \epsilon _{\mu \nu \lambda}  f_{\nu} (\bm{\nabla} f_{\lambda}), \label{sh2}
\end{eqnarray}
\begin{eqnarray}
\frac{\partial}{\partial t}  v_{\mu} &+& v_{\nu} \nabla _{\nu} v_{\mu} - \frac{\hbar ^{2}}{2M^{2}} \nabla _{\mu} \frac{\nabla ^{2}_{\nu} \sqrt{\rho}}{\rho}  \nonumber  \\ 
&+& \frac{\hbar^{2}}{4M^{2}\rho} \nabla _{\nu} \rho \biggl\{ (\nabla_{\mu}f_{\lambda})(\nabla_{\nu}f_{\lambda}) 
- f_{\lambda}(\nabla_{\mu} \nabla_{\nu} f_{\lambda})  \biggr\}  \nonumber  \\ 
&=& -\frac{1}{M}\biggl\{ c_{0}(\nabla _{\mu} \rho) + c_{1} f_{\nu} (\nabla _{\mu} \rho f_{\nu}) \biggr\} . \label{sh3}
\end{eqnarray}
In the next section, we use these equations to investigate SWT analytically. 

\section{Application of WWT theory to the spin hydrodynamic equation}
We apply  WWT theory \cite{wt1,wt2} to the spin hydrodynamic equation, obtaining two power laws in the transverse spin correlation function for the direct and inverse cascades, respectively. In this derivation, we make use of the previous result obtained in Ref. \cite{Lut}. 

First, we derive the equation of motion for fluctuations of the spin vector. Second, neglect of the spin-velocity and spin-density interactions is shown to make Eq. (\ref{sh1}) the Heisenberg ferromagnetic equation, and we find that our SWT is equivalent to the previous one \cite{Lut}. 
Finally, we derive the power laws for the transverse spin correlation function, which can be observed experimentally. 

\subsection{Approximation of the spin hydrodynamic equation}
In our SWT, there are weak fluctuations that cause the ferromagnetic ground state to deviate from being a fully spin-polarized state. To address these fluctuations, we decompose the spin vector, total density, and velocity field into their spatially averaged values and fluctuations as follows:
\begin{eqnarray}
\rho = \bar{\rho} + \delta \rho, 
\end{eqnarray}
\begin{eqnarray}
f_{\mu} = \bar{f}_{\mu} + \delta f_{\mu},  \label{spin_fluc}
\end{eqnarray}
\begin{eqnarray}
\bm{v} = \delta \bm{v}.
\end{eqnarray}
Here, the spatially averaged values are defined by $\bar{\rho}=\langle \rho \rangle _{\rm V}=N/V$ and $\bar{f}_{\mu}=\langle f_{\mu} \rangle _{\rm V}$, with the spatially averaged operation $\langle \cdots \rangle _{\rm V}= \int \cdots \hspace{1mm} dV/V$,  total particle number $N$, and  system volume $V$. These averaged quantities are independent of time owing to the symmetry of the Hamiltonian for the spin-1 spinor GP equation (\ref{GP}).
In our case, the ground state is assumed to be the fully spin-polarized state in the $z$ direction, having $\bar{f}_{\mu} = \delta_{\mu z}$.

Substituting these fluctuations into the spin hydrodynamic equations (\ref{sh0})--(\ref{sh2}) and retaining the spin-spin, spin-density, and spin-velocity interaction terms, we obtain
\begin{eqnarray}
\frac{\partial}{\partial t} \delta f_{\mu} = I_{\mu}^{\rm (s)} + I_{\mu}^{\rm (ss)} + I_{\mu}^{\rm (sd)}  + I_{\mu}^{\rm (sv)},
\end{eqnarray}
\begin{eqnarray}
I_{\mu}^{\rm (s)} = \frac{\hbar}{2M} \epsilon_{\mu\nu\lambda} \bar{f}_{\nu} \triangle \delta f_{\lambda},
\end{eqnarray}
\begin{eqnarray}
I_{\mu}^{\rm (ss)} = \frac{\hbar}{2M} \epsilon_{\mu\nu\lambda} \delta f_{\nu} \triangle \delta f_{\lambda}, 
\end{eqnarray}
\begin{eqnarray}
I_{\mu}^{\rm (sd)} = \Bigl[ \frac{\hbar}{2M\rho} \epsilon_{\mu\nu\lambda} f_{\nu}(\bm{\nabla} \delta f_{\lambda})  \cdot \bm{\nabla} \Bigl] \delta \rho,
\end{eqnarray}
\begin{eqnarray}
I_{\mu}^{\rm (sv)} = -(\delta \bm{v}\cdot \bm{\nabla})\delta f_{\mu},  
\end{eqnarray}
where each term $I_{\mu}^{\rm (a)}$ (a = s, ss, sd,  and sv) are the spin-linear, spin-spin interaction, spin-density interaction, and spin-velocity interaction terms. 

At present, we do not sufficiently understand which interactions are dominant. This issue should  depend on the initial state, interaction parameters, and methods of excitation of the system. However, in our SWT, we expect that the spin-spin interaction is stronger than other interactions because (I) the density fluctuation seems to be weak since the excitation to the system is not very strong and the inequality $c_{0} \gg |c_{1}|$ is satisfied, and (II) the velocity fluctuation is also expected to be weak since the spin and density fluctuations are weak. 
Hence, we may neglect the spin-density and spin-velocity interaction terms in SWT. 
However, this is speculation; therefore, in numerical calculations, we must confirm whether or not this situation is satisfied. 

Based on this speculation, we can neglect the spin-density and spin-velocity interactions, deriving the equation of motion for the spin vector $f_{\mu}$:
\begin{eqnarray}
\frac{\partial}{\partial t} f_{\mu} = \frac{\hbar}{2M } \epsilon_{\mu\nu\lambda} f_{\nu}  \triangle f_{\lambda}.
\label{spin_eq1}
\end{eqnarray}
Here, we use Eq. (\ref{spin_fluc}), changing the variable $\delta f_{\mu}$ to $f_{\mu}$. This equation is the same as the Heisenberg ferromagnetic or the Landau-Lifshitz equation, which was originally used in solid state physics \cite{anderson}.  

In the previous study \cite{Lut}  WWT theory was applied to Eq. (\ref{spin_eq1}) to study SWT in magnetic substances, allowing investigation of the distribution of the spin wave and finding the two power laws for the direct and inverse cascades. 

\subsection{Power laws in the spin-wave distribution  }
We briefly review  some of the results obtained in \cite{Lut}, wherein SWT in three-dimensional systems was theoretically studied by applying WWT theory to Eq. (\ref{spin_eq1}). For this application, it is necessary to transform Eq. (\ref{spin_eq1}) to the canonical form. Then, the new complex variable $a$ is usually introduced as follows \cite{wt1}:
\begin{eqnarray}
f_{+} = f_x + if_{y} = a\sqrt{2-a^{*}a},
\label{HP1}
\end{eqnarray}
\begin{eqnarray}
f_{-} = f_x - if_{y} = a^{*}\sqrt{2-a^{*}a},
\label{HP2}
\end{eqnarray}
\begin{eqnarray}
f_{z}  = 1-a^{*}a, 
\label{HP3}
\end{eqnarray}
which is the classical version of the Holstein-Primakoff transformation. These are the canonical variables obeying
\begin{eqnarray}
i\hbar \frac{\partial}{\partial t} a = \frac{\delta W}{\delta a^{*}}, 
\label{spin_eq2}
\end{eqnarray}
\begin{eqnarray}
W = \frac{\hbar^2}{4M} \int (\nabla_{\mu} f_{\nu})^2 dV. 
\label{spin_eq3}
\end{eqnarray}

In SWT, the amplitude of $|a|^2$ is much smaller than unity, so  we can expand $W$ with the power of the canonical variable $a$. Retaining the leading interaction between the spin waves, we obtain
\begin{eqnarray}
W = W_{0} + W_{1},  
\label{spin_eq4}
\end{eqnarray}
\begin{eqnarray}
W_{0} = \frac{\hbar^2}{2M} \int |\bm{\nabla}a|^2   dV, 
\label{spin_eq5}
\end{eqnarray}
\begin{eqnarray}
W_{1} = \frac{\hbar^2}{8M} \int \Bigl[ a^2(\bm{\nabla}a^*)^2 + (a^*)^2(\bm{\nabla}a)^2 \Bigl]  dV. 
\label{spin_eq6}
\end{eqnarray}

To consider the dynamics in  wave-number space, we perform the Fourier transformation on the canonical equation (\ref{spin_eq2}), obtaining
\begin{eqnarray}
i\hbar \frac{\partial}{\partial t} \hat{a}(\bm{k}) = \frac{\partial H}{\partial \hat{a}^{*}(\bm{k})}, 
\label{spin_eq7}
\end{eqnarray}
with the Fourier component $\hat{a} (\bm{k})=\mathcal{F}[a](\bm{k}) =  \int a(\bm{r}) {\rm exp} (-i \bm{k} \cdot \bm{r}) dV /V$. 
The Hamiltonian of the spin wave is given by
\begin{eqnarray}
H = H_{0} + H_{1},  
\label{spin_eq8}
\end{eqnarray}\begin{eqnarray}
H_{0} = \sum _{\bm{k}_1} \epsilon(k_1) |\hat{a}(\bm{k}_1)|^{2}, 
\label{spin_eq9}
\end{eqnarray}
\begin{eqnarray}
H_{1} =  \frac{1}{2} \sum _{\bm{k}_1,\bm{k}_2,\bm{k}_3,\bm{k}_4} W^{1,2}_{3,4} \delta^{1,2}_{3,4} \hat{a}^{*}(\bm{k}_1) \hat{a}^{*}(\bm{k}_2)\hat{a}(\bm{k}_3)\hat{a}(\bm{k}_4), 
\label{spin_eq10}
\end{eqnarray}
\begin{eqnarray}
W^{1,2}_{3,4}=-\frac{\hbar^{2}}{4M}\Bigl[ (\bm{k}_1\cdot\bm{k}_2) + (\bm{k}_3\cdot\bm{k}_4) \Bigl], 
\label{spin_eq11}
\end{eqnarray}
where $\delta^{1,2}_{3,4} =\delta(\bm{k}_1+\bm{k}_2-\bm{k}_3-\bm{k}_4)$ and $\epsilon (k)=\hbar ^2 k^2/2M$ are the Kronecker $\delta$ and the excitation energy for the spin wave, respectively. 

We focus on the distribution of the spin wave, which is defined by 
\begin{eqnarray}
n(\bm{k}) = \biggl( \frac{L}{2\pi} \biggl)^{d} \langle |\hat{a}(\bm{k})|^{2} \rangle _{\rm en}, 
\label{spin_dis}
\end{eqnarray}
with  system size $L$ and  system dimension $d=3$. Here the angular brackets $\langle  \cdots \rangle _{\rm en}$ mean the ensemble average. 
Using  WWT theory, we derive the kinetic equation of the spin wave as 
\begin{eqnarray}
\frac{\partial}{\partial t} n(\bm{k}) = \int \mathcal{R}^{k,1}_{2,3}~d\bm{k}_1 d\bm{k}_2 d\bm{k}_3,   \label{kin_equ1}
\end{eqnarray}
\begin{eqnarray}
\mathcal{R}^{k,1}_{2,3} &=& 4 \pi |W^{k,1}_{2,3}|^2  \delta _{\rm d}(\epsilon^{k,1}_{2,3}) {\delta_{\rm d}} ^{k,1}_{2,3} \nonumber \\
&\times&n(\bm{k}_1)n(\bm{k}_2)n(\bm{k}_3)n(\bm{k}) \nonumber \\  
&\times& \Biggl( \frac{1}{n(\bm{k})}+\frac{1}{n(\bm{k}_1)}-\frac{1}{n(\bm{k}_2)}-\frac{1}{n(\bm{k}_3)} \Biggl).   \label{kin_equ2}
\end{eqnarray}
Here, the abbreviations for the Dirac delta function  $\delta _{\rm d}$ are defined by ${\delta_{\rm d}} ^{k,1}_{2,3} = \delta_{\rm d} (\bm{k}+\bm{k}_1-\bm{k}_2-\bm{k}_3)$ and ${\delta_{\rm d}} (\epsilon ^{k,1}_{2,3}) = \delta_{\rm d} (\epsilon (k)+\epsilon (k_1)-\epsilon (k_2)-\epsilon (k_3))$. 

In the previous study \cite{Lut}  this equation was utilized to obtain two power laws in three-dimensional SWT:
\begin{eqnarray}
n(\bm{k}) \propto \left\{ \begin{array}{ll}
k^{-13/3} & ({\rm direct\hspace{1mm}cascade}), \\
k^{-11/3} & ({\rm inverse\hspace{1mm}cascade}). \\
\end{array} \right.
\end{eqnarray}
In SWT, the interaction between the spin waves is a four-wave interaction, so  the action of the spin wave is conserved.
Thus, the usual Fj\o rtoft argument \cite{wt1,wt2,Fjo} described in the Appendix leads to the direct cascade for the spin-wave energy and the inverse cascade for the action of the spin wave. 

We can easily extend this result to SWT in the two-dimensional system, obtaining 
\begin{eqnarray}
n(\bm{k}) \propto \left\{ \begin{array}{ll}
k^{-4/3-d} & ({\rm direct\hspace{1mm}cascade}) \\
k^{-2/3-d} & ({\rm inverse\hspace{1mm}cascade}), \\
\end{array} \right.
\label{spin_pow1}
\end{eqnarray}
with $d=2,3$. The locality of these power laws can be confirmed by the result of \cite{Kat}.

\subsection{Power laws in the transverse spin correlation function}
In experiments, the spin-density vector can be observed by the phase contrast imaging method \cite{KZ1,Ven,Sta}. 
Thus, we focus on the correlation function of the spin-density vector in SWT, defining the transverse spin correlation function: 
\begin{eqnarray}
C_{xy}^{\rm (s)}(k) = \frac{1}{\triangle k} \sum_{k-\triangle k/2 \leq |\bm{k}_1|<k+\triangle k/2} \langle |\hat{F}_{x}(\bm{k}_1)|^2+|\hat{F}_{y}(\bm{k}_1)|^2 \rangle_{\rm en}, 
\label{spin_corr}
\end{eqnarray}
with $\triangle k = 2 \pi /L$ and the Fourier component $\hat{F}_{\mu} (\bm{k})=\mathcal{F}[F_{\mu}](\bm{k})~(\mu=x,y,+,-)$. 
In our case, $\hat{F}_{+}(\bm{k}) \sim\sqrt{2} \bar{\rho} \hat{a}(\bm{k})$ is approximately satisfied, which leads to
\begin{eqnarray}
\hat{F}_{+}(\bm{k}) \hat{F}_{-}(\bm{k}) \propto |\hat{a}(\bm{k})|^2. 
\end{eqnarray}
Thus, we obtain
\begin{eqnarray}
C_{xy}^{\rm (s)}(k) \propto \left\{ \begin{array}{ll}
k^{-7/3} & ({\rm direct\hspace{1mm}cascade}) \\
k^{-5/3} & ({\rm inverse\hspace{1mm}cascade}). \\
\end{array} \right.
\label{spin_pow2}
\end{eqnarray}
In contrast to Eq. (\ref{spin_pow1}), these are independent of the system dimension $d$ because of the integration over the solid angle. 

\section{Numerical calculation for SWT} 
We now present our numerical results for SWT in a two-dimensional uniform system using the GP equation. 
Our numerical calculations are performed for the direct and inverse cascades in SWT, respectively.  
In the calculation for the inverse cascade, to induce the mode transfer from the high-wave-number to low-wave-number region, we use the initial state where the energy is injected into the high-wave-number region. In contrast, in the calculation for the direct cascade, the state into which the energy is injected in the low-wave-number region is adopted as the initial state. 

\subsection{SWT for the inverse cascade}

\subsubsection{1.Numerical condition and method}
Our calculation addresses the ferromagnetic spin-1 spinor BEC in a two-dimensional system whose system size $L \times L$ is $256\xi \times 256\xi$ with $\xi=\hbar/\sqrt{2Mc_0\bar{\rho}}$. We set the spatial resolution as $dx/\xi=1$. The time resolution is $dt/\tau=2\times10^{-3}$ with $\tau=\hbar/c_0\bar{\rho}$. The interaction parameters are taken to be $c_0>0$, $c_1<0$, and $|c_0/c_1|=20$. In this situation, we numerically solve the GP equation (\ref{GP}) by using the pseudo spectral method. 

We describe how to prepare the initial state. To confirm the inverse cascade, the energy should be injected into the initial state in the high-wave-number region. Then we adopt the following state as the initial state:
\begin{equation}
\begin{pmatrix}
\psi_{1}(\bm{r})  \\
\psi_{0}(\bm{r})  \\
\psi_{-1}(\bm{r})
\end{pmatrix}
= \sqrt{\bar{\rho}} {\rm e}^{i \phi(\bm{r}) }
\begin{pmatrix}
{\rm e}^{-i\alpha(\bm{r})} {\rm cos}^2 \frac{\beta(\bm{r})}{2}\\
\frac{1}{\sqrt{2}} {\rm sin } \beta(\bm{r}) \\
{\rm e}^{i\alpha(\bm{r})} {\rm sin}^2 \frac{\beta(\bm{r})}{2}
\end{pmatrix}
,
\label{ini1}
\end{equation}

\begin{eqnarray}
\mathcal{F}[\alpha](\bm{k}) =  p_1(R_1 + i R_2) {\rm exp} \Bigl[ - \{ (k \xi -0.4)/0.2 \}^2 \Bigl], 
\label{ini2}
\end{eqnarray}
\begin{eqnarray}
\mathcal{F}[\beta](\bm{k}) =  p_2(R_3 + i R_4) {\rm exp}\Bigl[ - \{ (k \xi -0.4)/0.2 \}^2 \Bigl], 
\label{ini3}
\end{eqnarray}
\begin{eqnarray}
\phi (\bm{r}) = \alpha (\bm{r}), 
\label{ini4}
\end{eqnarray}
where $R_i$ $(i=1$--4) are the random numbers in the range $[-0.5,0.5)$. The functions $\alpha$ and $\beta$ denote the azimuthal and elevation angles, respectively, for the spin-density vector, and $\phi$ is the phase of the wave function. Because of spin-gauge symmetry in the ferromagnetic state, $\phi$ also contains an angle related to the rotation around the spin-density-vector axis. The parameters $p_1$ and  $p_2$ are set to be $\langle f_z \rangle _{\rm V} \sim0.95$, ${\rm max} \hspace{1mm} [\alpha] \sim180$, and ${\rm min}  \hspace{1mm}  [\alpha] \sim -180$. 

Let us comment on Eq. (\ref{ini4}). In the ferromagnetic state, the superfluid velocity is expressed by 
\begin{eqnarray}
\bm{v} = \frac{\hbar}{M} \Bigl[ \bm{\nabla}\phi - {\rm cos}\beta \bm{\nabla} \alpha  \Bigl], 
\label{vel_ferro}
\end{eqnarray}
from which we can find that the gradients of $\alpha$ and $\phi$ induce the velocity field. 
Thus, if the relation of Eq. (\ref{ini4}) is not satisfied, the system has a large velocity field, which can cause  strong spin-density and spin-velocity interactions. This can disturb the assumption used to derive the power laws of Eq. (\ref{spin_pow2}). Hence, to reduce these interactions, we choose Eq. (\ref{ini4}). 

\subsubsection{2.Numerical results}

\begin{figure} [t]
\begin{center}
\includegraphics[width=85mm]{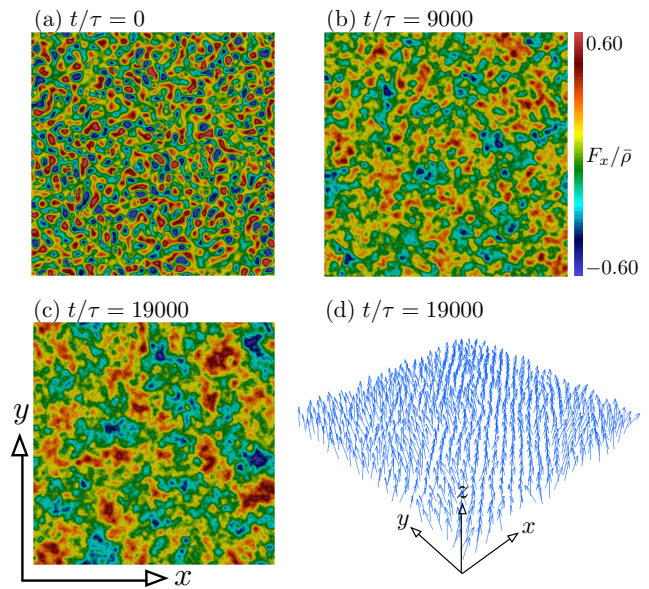}
\caption{(Color online)  Time development of the spin-density vector $\bm{F}$ in SWT for the inverse cascade.  We plot the spatial distribution of $F_x$ at $t/\tau=$ (a) 0, (b) 9000, and (c) 19000. (d) is the spatial profile of $\bm{F}$ corresponding to (c). The size of the figures is $ 256\xi \times 256\xi$. These figures show that the larger structure appears as time passes, which reflects the inverse cascade. }
\label{fig1}
\end{center}
\end{figure}

\begin{figure} [t]
\begin{center}
\includegraphics[width=82mm]{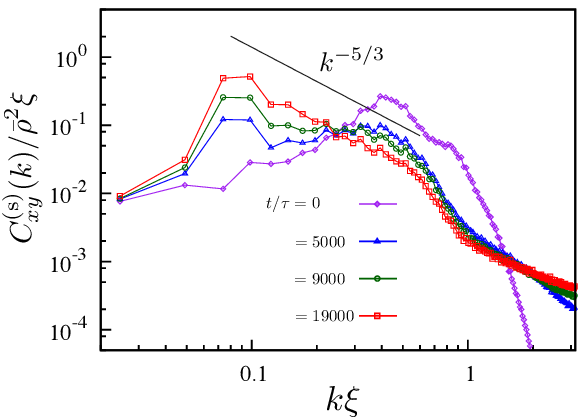}
\caption{(Color online)  Time development of the transverse spin correlation function in SWT for the inverse cascade. The black solid line in the range [0.08,0.6] exhibits a $k^{-5/3}$ power law. The spectra are averaged over five calculations with different initial noise components. }
\label{fig2}
\end{center}
\end{figure}

\begin{figure} [b]
\begin{center}
\includegraphics[width=90mm]{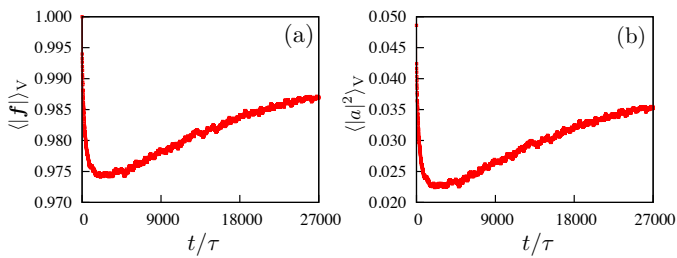}
\caption{(Color online) Time development of (a) the spatial average of the spin amplitude, $\langle |\bm{f}| \rangle _{\rm V}$, and (b) the spatial average of the canonical variable, $\langle |a|^2 \rangle _{\rm V}$, which satisfy approximations (i) and (ii).}
\label{fig3}
\end{center}
\end{figure}

\begin{figure} [t]
\begin{center}
\includegraphics[width=80mm]{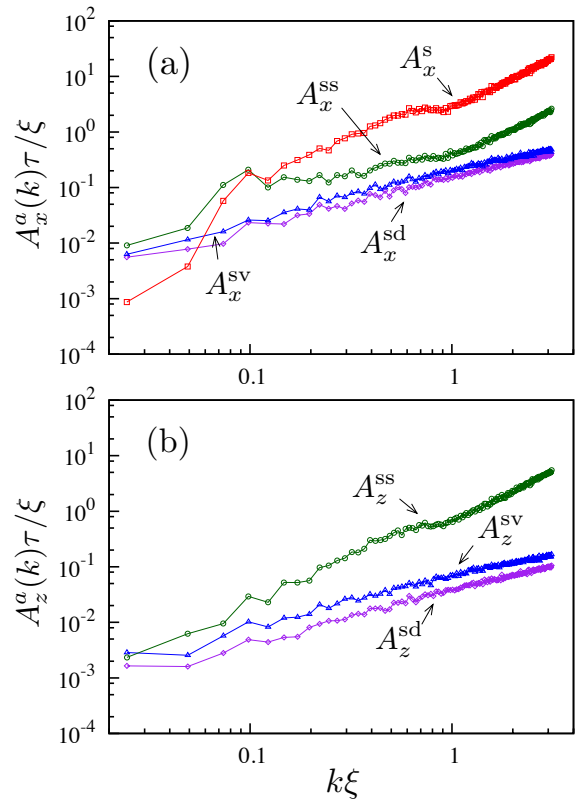}
\caption{(Color online) Wave-number dependence of $A^{\rm a} _{\mu}$ (a = s,\;ss,\;sd, and sv; $\mu=x,z$) at $t/\tau=19000$. From the rotational symmetry about the $z$ axis, the behavior of $A_{y}^{a}$ is the same as that of $A_{x}^{a}$. These graphs show that the spin-density and spin-velocity interaction terms are slightly smaller than the spin-spin interaction. }
\label{fig4}
\end{center}
\end{figure}

Figure \ref{fig1} shows the time development of the spin-density vector $\bm{F}$. Figure \ref{fig1}(a) is for the $x$ component of the spin-density vector in the initial state and shows that $F_x$ has the spatially finer structure corresponding to the energy injection in the high-wave-number region. As time passes, large-scale structure [Figs. \ref{fig1}(b) and (c)] appears, which reflects the inverse cascade. In Fig. \ref{fig1}(d), we plot the spin-density vector corresponding to Fig. \ref{fig1}(c), from which the spin-density vector is found to point in the $z$ direction and fluctuate around it. 

We numerically calculate the transverse spin correlation functions, showing their time development in Fig. \ref{fig2}. 
In the initial state, the correlation function has a large value in the wave-number region near $k\xi=0.4$, whose wavelength is comparable to the characteristic size of $F_{x}$ in Fig. \ref{fig1}(a). In the early stage of the dynamics, the inverse cascade leads to the growth of $C_{xy}^{\rm (s)}$ in the wave number region lower than $k\xi=0.4$, and the $-5/3$ power law appears only in the region near $k\xi=0.4$. As sufficient time passes, the scaling region with the $-5/3$ power law becomes wide and finally reaches $0.08 \alt k\xi \alt 0.6$. This behavior is consistent with Eq. (\ref{spin_pow2}) for the inverse cascade predicted by WWT theory. 

To investigate our SWT in detail, we check whether or not our approximations used to derive Eq. (\ref{spin_pow2}) are satisfied. The important approximations are (i) the assumption of a ferromagnetic state, (ii) the smallness of the canonical variables, and (iii) the weakness of the spin-density interaction $I^{\rm sd} _\mu$ and spin-velocity interaction $I^{\rm sv} _\mu$. 

To confirm (i), the amplitude $|\bm{f}|$ of the spin-density vector is calculated; it should be unity in the fully ferromagnetic state. Figure \ref{fig3}(a) shows the time development of its spatial average and indicates that the system almost becomes a ferromagnetic state.  For (ii), we calculate the amplitude $|a|^2$ of the canonical variables defined by Eqs. (\ref{HP1})--(\ref{HP3}), confirming that its spatially averaged value is much smaller than unity, as shown in Fig. \ref{fig3}(b). 
To check (iii), we define the following quantities: 
\begin{eqnarray}
A_{\mu}^{\rm a}(k) =  \frac{1}{\triangle k} \sum_{k-\triangle k/2 \leq |\bm{k}_1|<k+\triangle k/2} |\mathcal{F}[I^{\rm (a)}_{\mu}] (\bm{k}_1)|,   
\label{weak1}
\end{eqnarray}
with ${\rm a = s,\; ss,\; sd,\; and\;  sv}$ and $\mu =x,y,z$. 
Figure \ref{fig4} shows the numerical results for $A_{\mu}^{\rm a}(k)$ at $t/\tau=19000$ when the $-5/3$ power law appears over a wide interval. 
These figures show that the spin-density interaction $I^{\rm sd} _\mu$ and spin-velocity interaction $I^{\rm sv} _\mu$ are slightly weaker than the spin-spin interaction $I^{\rm ss} _\mu$ in the scaling region $0.08 \alt k\xi \alt 0.6$. In summary, our calculation basically satisfies approximations (i), (ii), and (iii). However, for (iii), the spin-density and spin-velocity interactions are not very weak.  At present, we do not sufficiently understand why the influence of density and velocity does not disturb the $-5/3$ power law.

In comparison with $A_{\mu}^{\rm s}(k)$, $A_{\mu}^{\rm ss}(k)$ is found to be weak in the region higher than $k\xi \sim 0.15$, which corresponds to the smallness of the canonical variable $a$. However, in the low-wave-number region $k\xi \alt 0.15$, the strength of the spin-spin interaction term is larger than that of the spin-linear term, so WWT becomes invalid. Actually, in this region, the $-5/3$ power law seems to be disturbed, as shown in Fig. \ref{fig2}. Thus, we conclude that the scaling region with the $-5/3$ power law is about $0.15 \alt k\xi \alt 0.60$. 

Finally, we describe the dynamics at $t/\tau > 19000$. The transferred mode accumulates in the low-wave-number region because any dissipation mechanism is not included in our numerical calculation. This can disturb the $-5/3$ power law. Actually, in the low-wave-number region, the correlation function is found to tend to deviate from the $-5/3$ power law when sufficient time has passed. 

\subsubsection{3.Dependence on initial conditions}
We describe how the $-5/3$ power law depends on the initial state. In our calculation, as long as we use an initial state with  energy injection in the wave-number region $0.2 \alt k\xi \alt 0.8$, the $-5/3$ power law appears. If we inject the energy into an initial state in the wave number region higher than $0.9/\xi$, the numerical precision is poor, so we cannot investigate the $-5/3$ power law in this case. 

\subsection{SWT for the direct cascade}

\subsubsection{1.Numerical conditions and method}
The numerical setting for the direct cascade is the same as that of the inverse cascade except for the system size, numerical resolution, and initial state. 
The system size is $64\xi \times 64\xi$ and the resolution is given by $dx/\xi=0.25$ and $dt/\tau=4\times10^{-3}$. For the initial state, in contrast to the inverse cascade, it is necessary for the energy to be injected into an initial state in the low-wave-number region. Then we use Eq. (\ref{ini1}) and the following angles as the initial state:
\begin{eqnarray}
\mathcal{F}[\alpha](\bm{k}) =  p_1(R_1 + i R_2) {\rm exp} \Bigl[ -(k \xi/0.2)^2  \Big], 
\label{ini5}
\end{eqnarray}
\begin{eqnarray}
\mathcal{F}[\beta](\bm{k}) =  p_2(R_3 + i R_4)  {\rm exp}\Bigl[ -(k \xi/0.2)^2 \Bigl], 
\label{ini6}
\end{eqnarray}
\begin{eqnarray}
\phi (\bm{r}) = \alpha (\bm{r}).
\label{ini7}
\end{eqnarray}
The parameters $p_1$ and  $p_2$ are set to be $\langle f_z \rangle _{\rm V} \sim0.95$, ${\rm max} \hspace{1mm} [\alpha] \sim180$, and ${\rm min}  \hspace{1mm}  [\alpha] \sim -180$. 

\begin{figure} [t]
\begin{center}
\includegraphics[width=85mm]{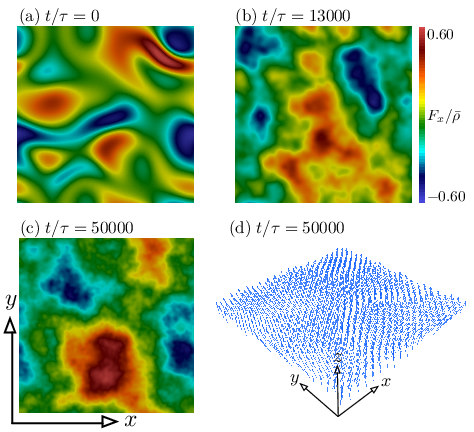}
\caption{(Color online) Time development of the spin-density vector $\bm{F}$ in SWT for the direct cascade.  We plot the spatial distribution of $F_x$ at $t/\tau=$ (a) 0, (b) 13000, and (c) 50000. (d) is the spatial profile of $\bm{F}$ corresponding to (c). The size of the figures is $64\xi \times 64 \xi$. These figures show that the finer structure appears as time passes, which reflects the direct cascade. }
\label{fig5}
\end{center}
\end{figure}

\begin{figure} [b]
\begin{center}
\includegraphics[width=82mm]{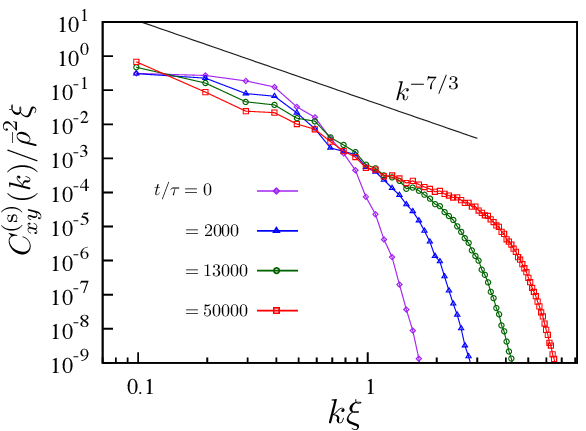}
\caption{(Color online)  Time development of the transverse spin correlation function in SWT for the direct cascade. The black solid line in the range [0.1,3.0] exhibits a $k^{-7/3}$ power law.  The spectra are averaged over five calculations with different initial noise components. }
\label{fig6}
\end{center}
\end{figure}

\begin{figure} [t]
\begin{center}
\includegraphics[width=90mm]{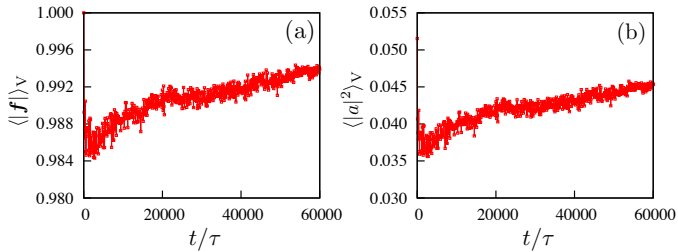}
\caption{(Color online) Time development of (a) the spatial average of the spin amplitude, $\langle |\bm{f}| \rangle _{\rm V}$, and (b) the spatial average of the canonical variable, $\langle |a|^2 \rangle _{\rm V}$. These graphs show that the system is almost in a fully ferromagnetic state with a small spin-wave amplitude, which satisfies  approximations (i) and (ii).} 
\label{fig7}
\end{center}
\end{figure}

\begin{figure} [t]
\begin{center}
\includegraphics[width=80mm]{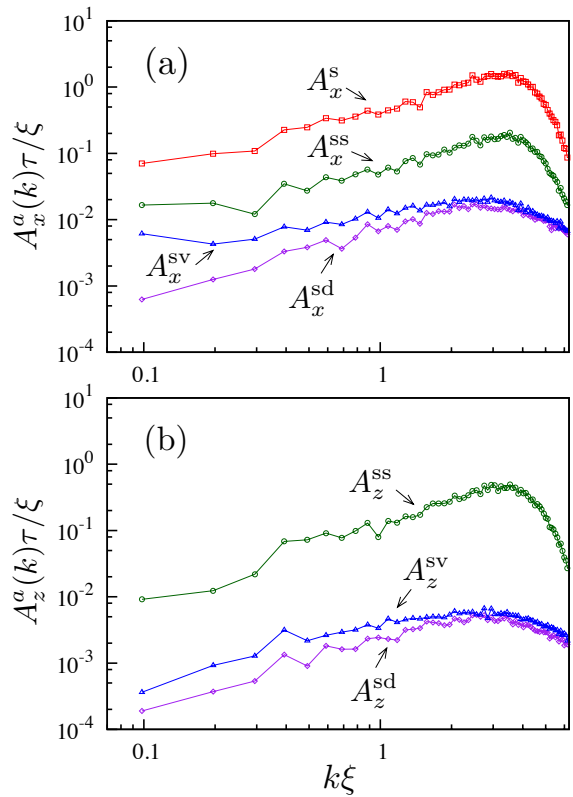}
\caption{(Color online) Wave-number dependence of $A^{\rm a} _{\mu}$ (a = s, ss, sd, and sv; $\mu=x,z$) at $t/\tau=50000$. From the rotational symmetry about the $z$ axis, the behavior of $A_{y}^{a}$ is the same as that of $A_{x}^{a}$. These graphs show that the spin-density and spin-velocity interaction terms are weaker than the spin-spin interaction. }
\label{fig8}
\end{center}
\end{figure}

\subsubsection{2.Numerical results}
Figure \ref{fig5} shows the time development of the spin-density vector $\bm{F}$. As shown in Fig. \ref{fig5}(a), the profile of $F_x$ in the initial state has a spatially large structure, which reflects the energy injection into the low-wave-number region. As  time passes, finer structure appears in Figs. \ref{fig5}(b) and \ref{fig5}(c), which is caused by the nonlinear terms because these terms generate interactions between different modes in the wave-number space. In Fig. \ref{fig5}(d), we plot the spin-density vector corresponding to Fig. \ref{fig5}(c), from which $f_{z}$ is found to have values near unity. 

As in the case for the inverse cascade, we numerically calculate the transverse spin correlation function. Figure \ref{fig6} shows the time development of this correlation function. In the initial state, the correlation function has a large value only in the low-wave-number region, which corresponds to  energy injection into the low-wave-number region. In the early stage of the dynamics, the spin wave in the low-wave-number region is transferred to the high-wave-number region, increasing the high-wave-number component of $C_{xy}^{\rm (s)}$, which reflects the direct cascade of the spin-wave energy. As time passes, the $-7/3$ power law appears in the region $1.0 \alt k\xi \alt 3.0$. Although the scaling region is not wide, this behavior is consistent with Eq. (\ref{spin_pow2}) for the direct cascade predicted by WWT theory. 

We note the deviation from the $-7/3$ power law in the low-wave-number region  $k\xi \alt 1.0$ in Fig. \ref{fig6}. We suspect that this may be caused by the inverse cascade. As described in the Appendix, when the spin-wave energy is transferred into the high-wave-number region, the spin-wave action is simultaneously transferred into the low-wave-number region. However, in the initial state, the wave action is sufficiently accumulated in the low-wave-number region, so that the inverse cascade may be suppressed. Thus, this obstruction of the inverse cascade may lead to the deviation from the $-7/3$ power law in the low-wave-number region.  

In the same way as for the inverse cascade, we also confirm the validity of the three approximations used for the derivation of Eq. (\ref{spin_pow2}): (i) the assumption of  a ferromagnetic state, (ii) the smallness of canonical variables, and (iii) the weakness of the spin-density interaction $I^{\rm sd} _\mu$ and spin-velocity interaction $I^{\rm sv} _\mu$. Figures \ref{fig7}(a) and \ref{fig7}(b) obviously show the validity of (i) and (ii). For (iii), from Fig. \ref{fig8}, we find that the spin-density interaction and spin-velocity interaction are weaker than the spin-spin interaction, so  approximation (iii) is satisfied. Also, the spin-linear term is larger than the spin-spin interaction, which means that WWT is realized. 

Finally, we describe the dynamics at $t/\tau > 50000$. In our numerical calculation, we do not add any dissipation terms to Eq. (\ref{GP}), so that the transferred energy accumulates in the high-wave-number region, which disturbs the $-7/3$ power law. Actually, we numerically confirm that, in the high-wave-number region, the correlation function tends to deviate from the $-7/3$ power law when sufficient time has passed. 

\subsubsection{3.Dependence on initial conditions}
We discuss how the $-7/3$ power law depends on the initial state. When we use the larger system size and prepare the initial state with energy injection into the low-wave-number region, the direct cascade does not appear clearly. We expect that, in this case, because the wave number of the excited spin wave is too low, it takes much time to transfer the mode from the low-wave-number to the high-wave-number region. Thus, to see the $-7/3$ power law for the direct cascade in numerical calculations, it seems to be necessary to prepare the situation described in Sec. IV-B.

\section{Comments on experimental observation of power laws in SWT}
We discuss the possibility of observing the $-7/3$ and $-5/3$ power laws for direct and inverse cascades experimentally. In atomic BECs, in contrast to solid sate physics, the spatial distribution of the spin-density vector is observable \cite{KZ1,Sta,Ven}. Thus, these power laws may be observed too. However, as noted in the following discussion, the experimental observation of Eq. (\ref{spin_pow2}) requires further investigation for SWT. 

Essential issues for observing the power laws in SWT are (A) a method to generate SWT, (B) the effect of inhomogeneity induced by the trapping potential, and (C) the establishment of a wide scaling region for the power laws. 

First, we discuss the issue (A). In our numerical calculation, we artificially prepare the initial state by using a random number, but experiments must use some realistic method to excite the system. For the observation of the $-5/3$ power law, it is necessary to excite the spin wave in the high-wave-number region. In this case, a candidate for this excitation method would be oscillation of a localized obstacle such as a Gaussian potential with small radius. On the other hand, as for the $-7/3$ power law, we must excite the spin wave in the low-wave-number region. Thus, candidates for the excitation methods are considered to be  the oscillation of the trapping potential or the oscillation of a localized obstacle with a large radius. In both cases, the wave number and amplitude of the spin wave excited should  depend on the frequency and amplitude of the oscillation of the obstacle or the trapping potential, so  detailed investigation of what is caused by these excitations is required. 

Second, the issue (B) is discussed. In experiments on cold atomic gases, the atoms composing the system are captured by a trapping potential such as a harmonic potential, which generates an inhomogeneous density profile. Thus, this inhomogeneity may disturb the power laws of Eq. (\ref{spin_pow2}) because these are theoretically derived in a uniform system without any trapping potential as shown in Sec. III. Currently, we expect that, if the system size, e.g., the Thomas-Fermi radius, is larger than the scaling region with power laws, the influence of the inhomogeneity on these power laws may be weak. To confirm this speculation, a numerical calculation for SWT in the trapping potential is needed. 

Third, we argue the issue (C). In the usual experiments, the spatial resolution for the phase-contrast imaging \cite{KZ1,Ven} is of the order of $1~{\rm \mu m}$, so that it should be much smaller than the length scale corresponding to the scaling region with power laws.  In our numerical calculation, we use the relation $\lambda = 2\pi /k$ and the typical coherence length $\xi = 0.5~{\rm \mu m}$, finding that thelength scale for the inverse cascade is about $5.2~{\rm \mu m} \alt \lambda \alt 20.9~{\rm \mu m}$ corresponding to the scaling region $0.15 \alt k\xi \alt 0.6$, and that for the direct cascade is about $1.0~{\rm \mu m} \alt \lambda \alt 3.1~{\rm \mu m}$ corresponding to $1 \alt k\xi \alt 3$. Therefore, the power law for the inverse cascade may be experimentally observable, but that for the direct cascade may be difficult to observe. 

Finally, we comment on the interaction parameter in our numerical calculation. The typical experiment on ferromagnetic spin-1 spinor BECs uses $^{87}{\rm Rb}$ with the interaction parameter $|c_0/c_1| \sim 200$. However, in the numerical calculation, we use $|c_0/c_1|=20$ since the ferromagnetic state is hard to break. Thus, to confirm whether or not these power laws appear in experiments, we must perform a numerical calculation supposing the realistic situation. 

In summary, for simplicity, our numerical calculation addresses a spin-1 spinor BEC in an unrealistic situation, so that we cannot conclude whether or not the power laws in SWT are observable. Thus, to clarify the possibility of experimental observation for the power laws in SWT, a detailed numerical calculation assuming the experimental situation is required. 

\section{Conclusion}
We have analytically and numerically studied  SWT in a uniform ferromagnetic spin-1 spinor BEC at zero temperature by using the spinor GP equations. We have derived the $-7/3$ and $-5/3$ power laws for the direct and inverse cascades in the transverse spin correlation function by using 
the previous result \cite{Lut}. Our numerical calculation of the spinor GP equation has yielded a numerical result consistent with these power exponents, although the scaling region with the power laws is not very wide. Also, we checked whether or not the approximations used to derive the power laws were valid. Finally, we discussed the experimental observation of the power laws. 

\section*{ACKNOWLEDGMENTS}
The authors are grateful to Y. Aoki for fruitful discussion. K. F.  was supported by a Grant-in-Aid for JSPS Fellows
(Grant No. 262524). M. T.  was supported by JSPS KAKENHI Grant No. 26400366 and MEXT KAKENHI “Fluctuation $\&$ Structure” Grant No. 26103526.

\renewcommand{\theequation}{A\arabic{equation}}
\setcounter{equation}{0}
\section*{Appendix : Fj\o rtoft argument}

\begin{figure} [b]
\begin{center}
\includegraphics[width=85mm]{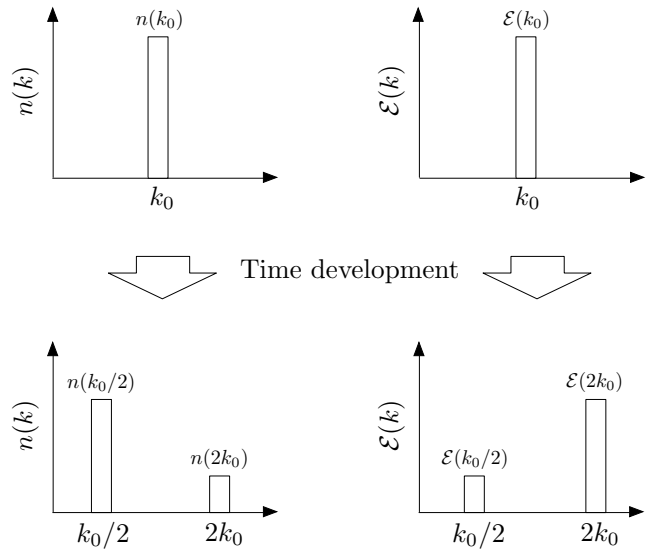}
\caption{Fj\o rtoft argument. The two conserved quantities restrict the dynamics, leading to the direct and inverse cascades (see text).}
\label{fig9}
\end{center}
\end{figure}

We describes the Fj\o rtoft argument \cite{wt1,wt2,Fjo}. In SWT, there are two conserved quantities defined by
\begin{eqnarray}
E = \sum _{\bm k} \mathcal{E}(k) = \sum _{\bm k} \epsilon (k) n(k) 
\end{eqnarray}
and
\begin{eqnarray}
N = \sum _{\bm k}  n(k), 
\end{eqnarray}
which are the spin-wave energy $E$ and action $N$. The dispersion relation $\epsilon (k)$ is given by $\hbar^2k^2 /2M$. 
Let us consider the situation shown in Fig. \ref{fig9}, where only a spin wave with a wave number $k_0$ is initially excited. After development in time, the spin wave is supposed to be redistributed into two spin waves with two wave numbers $k_0 /2$ and $2k_0$.  Through this dynamics, Eqs. (A1) and (A2) are independent of time, which leads to 
\begin{eqnarray}
\mathcal{E}(k_0) = \mathcal{E}(k_0/2) + \mathcal{E}(2k_0), 
\end{eqnarray}
\begin{eqnarray}
n(k_0) = n(k_0/2) + n(2k_0). 
\end{eqnarray}
We solve these coupled equations, obtaining
\begin{eqnarray}
n(k_0/2) = \frac{2}{5}n(k_0), 
\end{eqnarray}
\begin{eqnarray}
n(2k_0) = \frac{1}{5}n(k_0), 
\end{eqnarray}
\begin{eqnarray}
\mathcal{E}(k_0/2) = \frac{1}{5} \mathcal{E}(k_0), 
\end{eqnarray}
\begin{eqnarray}
\mathcal{E}(2k_0) = \frac{4}{5} \mathcal{E}(k_0). 
\end{eqnarray}
This result shows that the spin-wave energy is transferred from the low- to the high-wave-number region and vice versa  for the spin-wave action.  
Thus, direct and inverse cascades can occur in SWT because of the existence of the two conserved quantities. 
This discussion is called the Fj\o rtoft argument.

\end{document}